\documentstyle[11pt,newpasp,twoside,epsf]{article}
\def\edcomment#1{\iffalse\marginpar{\raggedright\sl#1\/}\else\relax\fi}
\reversemarginpar

\begin{document}
%\title{Diquark condensates and the magnetic field of pulsars}
\title{Magnetic field of pulsars with superconducting quark core}
\author{David Blaschke}
\affil{Fachbereich Physik, Universit\"at Rostock,
	Universit\"atsplatz 1, D--18051 Rostock, Germany}
\author{David M. Sedrakian and Karen M. Shahabasyan}
\affil{Physics Department, Yerevan State University, Alex
	Manoogian Str. 1, 375025 Yerevan, Armenia}

\begin{abstract}
Within recent nonperturbative approaches to the effective quark 
interaction the diquark condensate forms a superconductor of second kind. 
Therefore the magnetic field will not be expelled from the
superconducting quark core in accordance with observational data which 
indicate that life times of pulsar magnetic fields exceed $10^7$ years.   
\end{abstract}

%\section{Introduction}

The physical properties of pulsars can constrain 
our hypotheses about the state of matter at high densities. 
For example, Bailin and Love (1984) have suggested that the magnetic field of 
pulsars should be expelled from the superconducting interior of the star due 
to the Meissner effect and decay subsequently within $\approx 10^4$ years.
If their arguments would hold in general, the observation of lifetimes 
as large as $10^7$ years (Makashima 1992) would exclude the occurence of an 
extended superconducting quark matter phase in pulsars.  
For their estimate, they assumed a homogeneous magnetic 
field and used a perturbative quark interaction which results in a
very small pairing gap. 
Since both assumptions seem not to be valid in general, we perform a 
reinvestigation of this question.

Recently there has been excitement (Wilczek 1999) about the observation 
that in chiral quark models with nonperturbative 4-point interactions 
the anomalous quark pair amplitudes in the color antitriplet channel can be 
very large of the order $\approx 100~ {\rm MeV}$.
In Fig. 1 (a) we show the solution of the corresponding diquark gap 
equation for a BCS-type quark-quark interaction model 
(Berges \& Rajagopal 1999)
and the corresponding Ginzburg-Landau Parameter $\kappa$.
Quark matter with a diquark condensate appears as
a superconductor of second kind into which the magnetic field $H$ can 
penetrate by forming quantized vortex lines provided $H_{c1}< H <H_{c2}$ where 
$H_{c1}=1.8\cdot10^{16}$ G (Blaschke, Sedrakian \& Shahabasyan 1999).

It is generally accepted that neutrons and protons in the ``npe''-phase 
are superfluid.
While the neutrons take part in the rotation, forming a lattice of quantized 
vortex lines, the superconducting protons will be entrained by the neutrons 
(Sedrakian \& Shahabasian 1980) and form inside the neutron vortex a magnetic 
field strength $H(r)$ which
acts as an external field for the non-entrained protons. This entails the 
formation of a cluster of proton vortices with fluxes $\Phi_0$
in a region with the radius $\delta_n=10^{-5}$ cm around the axis of the 
neutron vortex.
The mean magnetic induction within the cluster 
reaches values of $4 \cdot 10^{14}$ G (Sedrakian \& Sedrakian 1995).
The magnetic field $H(r)$ generates quark vortex clusters with a radius 
$\delta_q=4.3 \cdot 10^{-7}$ cm. 
Since $\delta_q$ is by two orders of magnitude smaller than 
$\delta_n$, the mean magnetic induction in the clusters of quark vortex lines 
increases to a value of the order of $ 10^{18}$ G, see Fig. 1 (b).

\begin{figure}[ht]
\plottwo{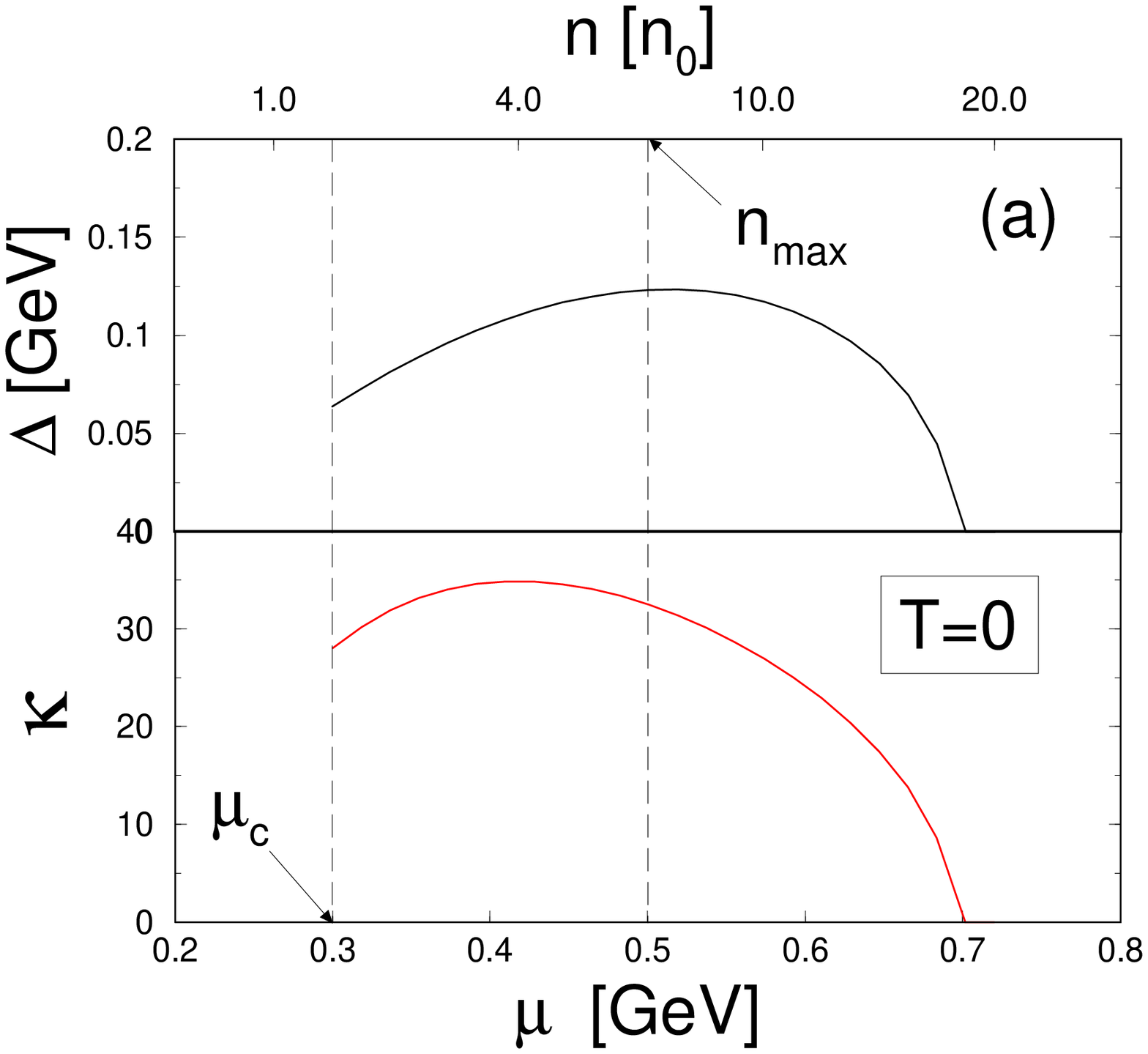}{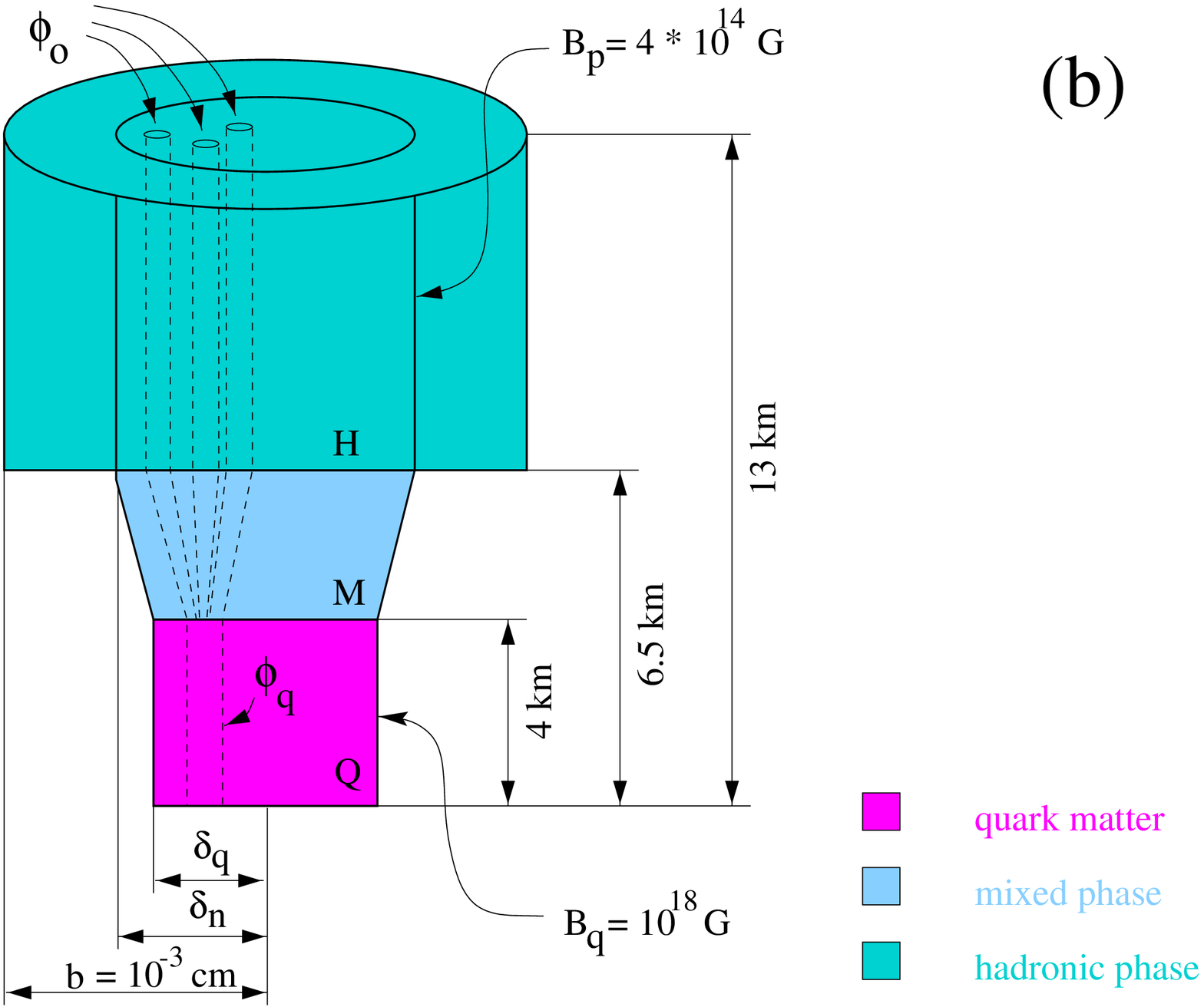}
\caption{(a) Diquark energy gap $\Delta$ and corresponding
Ginzburg-Landau parameter $\kappa$ vs. chemical potential $\mu$  
(resp. density n) for a BCS-type quark interaction.
(b) Magnetic field structure in the interior of a  hybrid 
star with $M=1.4 M_{\odot}$;
\label{vortex}}
\end{figure}
The clusters of quark vortex lines which appear due to the entrainment effect
in the ``npe''-phase will interact with those which are formed by the initial 
magnetic field (fossil field). 
Due to this interaction quark vortex
lines will not be expelled from the quark core of the star within a time scale
of $\tau=10^4$ years as suggested in (Bailin \& Love 1984).
This is the basic finding of our paper 
(Blaschke, Sedrakian \& Shahabasyan 1999).

{\acknowledgements}
K.M.S. acknowledges IAU support for the participation at the Conference.
This work has been supported in part by the Volkswagen Stiftung under
grant no.\ I/71 226.

\end{document}